\documentclass[11pt]{article}
\pdfoutput=1

\usepackage{amsfonts}
\usepackage{amsmath}
\usepackage{amssymb}
\usepackage{mathrsfs}
\usepackage[english]{babel}
\usepackage[T1]{fontenc}
\usepackage[latin1]{inputenc}
\usepackage{graphicx}

\numberwithin{equation}{section} 
\def\g{\gamma}
\def\G{\Gamma}
\def\d{\delta}

\def\ep{\epsilon}
\def\vep{\varepsilon}
\def\z{\zeta}
\def\e{\eta}
\def\th{\theta}

\def\la{\lambda}
\def\La{\Lambda}
\def\m{\mu}
\def\n{\nu}

\def\S{\Sigma}
\def\t{\tau}
\def\f{\phi}

\def\ps{\psi}
\def\o{\omega}

\providecommand{\abs}[1]{\lvert#1\rvert}

\newcommand{\emn}{\eta_{\mu \nu}}
\newcommand{\de}{\partial}
\newcommand{\dem}{\de_{\m}}

\newcommand{\p}{\prime}

\newcommand{\beq}{\begin{equation}}
\newcommand{\eeq}{\end{equation}}
\newcommand{\nn}{\nonumber}

\newcommand{\vecx}{\vec{x}}
\newcommand{\veck}{\vec{k}}
\newcommand{\vecp}{\vec{p}}

\newcommand{\vecP}{\vec{P}}

\newcommand{\mcal}[1]{\mathcal{#1}}
\newcommand{\mscr}[1]{\mathscr{#1}}

\begin{document}

\date{\mbox{}}

\title{
\vspace{-2.0cm}
\vspace{2.0cm}
{\bf \huge Classical and quantum ghosts}
 \\[8mm]
}

\author{
Fulvio Sbis\`a$^{1,2}$\thanks{fulvio.sbisa@gmail.com , fulvio.sbisa@port.ac.uk}
\\[8mm]
\normalsize\it
$^1$ Institute of Cosmology \& Gravitation, University of Portsmouth,\\
\normalsize\it Dennis Sciama Building, Portsmouth, PO1 3FX, United Kingdom\vspace{.5cm} \\
\normalsize\it
$^2\,$ Dipartimento di Fisica dell'Universit\`a di Milano\\
       {\it Via Celoria 16, I-20133 Milano} \\
       and \\
\normalsize\it
       INFN, Sezione di Milano, \\
       {\it Via Celoria 16, I-20133 Milano}\vspace{.5cm} \\[.3em]
}

\maketitle

\setcounter{page}{1}
\thispagestyle{empty}

\begin{abstract}

\noindent The aim of these notes is to provide a self-contained review of why it is generically a problem when a solution of a theory possesses ghost f\mbox{}ields among the perturbation modes. We def\mbox{}ine what a ghost f\mbox{}ield is and we show that its presence is associated to a classical instability whenever the ghost f\mbox{}ield interacts with standard f\mbox{}ields. We then show that the instability is more severe at quantum level, and that perturbative ghosts can exist only in low energy ef\mbox{}fective theories. However, if we don't consider very ad-hoc choices, compatibility with observational constraints implies that low energy ef\mbox{}fective ghosts can exist only at the price of giving up Lorentz-invariance or locality above the cut-of\mbox{}f, in which case the cut-of\mbox{}f has to be much lower that the energy scales we currently probe in particle colliders. We also comment on the possible role of extra degrees of freedom which break Lorentz-invariance spontaneously.

\end{abstract}

\smallskip

\section{Introduction}

During the last 15 years, ghost fields have raised considerable interest both from a phenomenological and a theoretical point of view. Despite this class of fields plays a role in several areas of physics (e.g.~the Faddeev-Popov ghosts in non-Abelian gauge theories \cite{Faddeev:1967}), the discovery of the cosmological late time acceleration has revived the interest in studying theories which contain ghosts. Regarding the dark energy paradigm, it has been proposed that dark energy could actually be a ``ghost condensate'' \cite{ArkaniHamed:2003uy} (see also \cite{ArkaniHamed:2003uz}), and from another point of view the possibility that the dark energy equation of state parameter $w$ be smaller than $-1$ has led to the proposal of phantom matter \cite{Caldwell:1999ew}, which can be realized by a scalar field with negative kinetic energy (see \cite{Caldwell:1999ew, Carroll:2003st} and references therein). On the other hand, concerning the modified gravity approach, modifying General Relativity (GR) in general introduces additional degrees of freedom which quite often are ghost fields, at least around some background solutions (e.g. the Boulware-Deser mode in massive gravity \cite{Boulware:1973my} and the bending mode in the DGP model around self-accelerating backgrounds, see \cite{Koyama:2007za} for a review). Indeed, only recently the problem of avoiding the presence of the Boulware-Deser ghost has been solved, with the formulation of a class of ghost-free and non-linear massive gravity \cite{deRham:2010kj, Hassan:2011hr} and bi--gravity theories \cite{Hassan:2011zd}. Furthermore, including higher derivatives in a theory generically leads to ghost instabilities, due to the Ostrogradski theorem \cite{Ostrogradski1850, Woodard:2006nt}. Unfortunately, the presence of ghosts in a theory is associated with a severe instability of the system, both at classical and quantum level, whose relevance is sometimes overlooked\footnote{The Faddeev-Popov ghosts are however harmless since they never appear as final states but only in internal loops.}. In these notes we want to give a self-contained and up-to-date discussion of the instability associated with ghost fields and stress its importance.

Let's consider for simplicity the following Lagrangian density for a free relativistic scalar field $\f$ in a 4D Minkowski spacetime (indices are raised/lowered with the flat metric $\e^{\m\n}$/$\e_{\m\n}$)
\beq
\label{ghostLagrangianfreeapp}
\mscr{L} = - \frac{\ep}{2} \, \dem \f \, \de^{\m} \f - \frac{\vep}{2} \, m^2 \f^2
\eeq
where $\ep = \pm 1$ and $\vep = \pm 1$ (note that we use the ``mostly plus'' convention for the signature, i.e. $\emn = \textrm{diag} ( -1,+1,+1,+1 )$). The momentum conjugated to $\f$ is defined by
\beq
\pi_{\f} \equiv \frac{\de \mscr{L}}{\de \dot{\f}} = \ep \dot{\f}  \quad ,
\eeq
and performing the Legendre transform with respect to $\dot{\f}$ (here an overdot indicates a time derivative) we obtain the Hamiltonian density
\beq
\label{ghostHamiltonianfreeapp}
\mscr{H} = \ep \, \Big( \frac{1}{2} \, \dot{\f}^2 + \frac{1}{2} \, \big( \vec{\nabla} \f \big)^{\! 2} \Big) + \frac{\vep}{2} \, m^2 \f^2
\eeq
in terms of which the Hamiltonian is defined as
\beq
H \equiv \int_{\mathbb{R}^3} \! d^3 x \,\, \mscr{H}[\f,\dot{\f}] \quad .
\eeq
In the $\ep =\vep = +1$ case the Hamiltonian is positive semi-definite and therefore bounded from below, while in the $\ep =\vep = -1$ case the Hamiltonian is negative semi-definite and therefore bounded from above. In the case $\ep = - \vep$, the Hamiltonian is indefinite and so it is not bounded either from below or from above. The field $\f$ is called a \emph{ghost field} if $\ep = \vep = -1$, while is called a \emph{tachyon field} if $\ep = +1$ and $\vep = -1$; finally, it is called a \emph{tachyonic ghost} if $\ep = -1$ and $\vep = +1$. Despite these definitions were given for a relativistic scalar field, it is straightforward to extend them to more general cases: a ghost field is defined as a field which has negative kinetic energy. If the Lagrangian density is not Lorentz-invariant, the part of the kinetic term which decides if the field is a ghost or not is the one which contains the time derivative of the field (the ``velocity'' of the field), or the conjugate momentum in the Hamiltonian formulation.

In these notes we consider a Lagrangian theory of $N$ fields $\{ \Psi_i \}_{i = 1, \ldots, N}$ defined on the (4-dimensional) Minkowski spacetime and indicate with $\{ \bar{\Psi}_i \}_{i = 1, \ldots, N}$ a solution of the equations of motion. In complete generality, we can re-express the theory in terms of the fluctuation fields $\psi_i \equiv \Psi_i - \bar{\Psi}_i$: we say that the theory has a perturbative ghost (around the solution $\{ \bar{\Psi}_i \}$) if one of the fluctuation fields $\psi_i$ is a ghost (in which case for clarity we will indicate it with $\phi$). The solution $\{ \bar{\Psi}_i \}$ of the original theory corresponds to the ``vacuum'' solution $\psi_i = \phi_i = 0$ of the ordinary/ghost perturbation fields. In the following, when we say that the vacuum is rendered unstable by the presence of ghosts we implicitly mean to say that the configuration $\bar{\Psi}_i$ is made unstable by the presence of perturbative ghosts. The particular case where the ghost is one of the ``fundamental'' fields $\{\Psi_i \}_i$ is included in this analysis as the case $\{ \bar{\Psi}_i \}_{i} = 0$.

\section{Ghosts at classical level}

A Hamiltonian which is unbounded from below is usually associated with instabilities of the system. However, if a ghost field $\f$ is free, the system is actually stable since the energy is conserved by time evolution, independently of its sign. In fact, at classical level an overall sign (or more in general a constant) in front of the Lagrangian density of the system has no influence at all, since it does not appear in the equations of motion. Therefore, at classical level, the theory described by the Lagrangian density (\ref{ghostLagrangianfreeapp}) corresponding to $\ep = \vep = +1$ is completely equivalent to theory described by the Lagrangian density corresponding to $\ep = \vep = -1$, and is defined in both cases by the equation of motion (the Klein-Gordon equation)
\beq
\big( \Box - m^2 \big) \f = 0
\eeq
where $\Box$ is the D'Alembert operator. If we consider the following Fourier decomposition
\beq
\f(\vecx, t) = \int_{\mathbb{R}^3} \! \frac{d^{3}p}{(2 \pi)^3} \,\, \f_{\vecp} (t) \,\, e^{i \vecp \cdot \vecx}
\eeq
we have that every mode is decoupled and obeys the equation
\beq
\ddot{\f}_{\vecp} (t) = - (m^2 + \vec{p}^{\,\,2}) \, \f_{\vecp} (t)  \quad ,
\eeq
which has only oscillatory solutions of frequency $\o(\vecp) = \sqrt{m^2 + \vec{p}^{\,\,2}}$. Since the plane waves of the Fourier expansion are orthonormal functions, a small perturbation\footnote{We define a perturbation $f(\vecx)$ to be small (respectively, big) if $\int d^3x \, f^2 \ll 1$ (respectively, $\gg 1$).} at $t = t_0$ from the configuration $\f = 0$ has small Fourier coefficients $\f_{\vecp} (t_0)$, and the oscillatory behaviour ensures that the perturbation remains small at all time. Therefore, the trivial configuration $\f(\vecx , t) = 0$ is stable both in the case $\ep = +1$ and in the case $\ep = -1$. Note instead that, if $\ep = -\vep$, the frequency $\o(\vecp) = \sqrt{\vec{p}^{\,\,2} - m^2}$ becomes imaginary for modes characterized by $\vec{p}^{\,\, 2} < m^2$ and so these mode can grow exponentially, signalling an instability.

However, the situation changes if a (classical) ghost field interacts with a (classical) non-ghost field. Consider in fact the following Lagrangian density for the relativistic scalar fields $\phi$ and $\psi$
\beq
\label{ghostLagrangianinteractingapp}
\mscr{L} = - \frac{\ep}{2} \, \dem \phi \, \de^{\m} \phi  - \frac{\ep}{2} \, m_{\f}^2 \f^2 - \frac{1}{2} \, \dem \psi \, \de^{\m} \psi - \frac{1}{2} \, m_{\psi}^2 \psi^2 - V_{\textup{int}} \big(\phi, \psi \big) \quad ,
\eeq
where we assume that the potential does not contain derivative interaction terms, is analytic in $\phi, \psi$ and that the configuration $\f = \ps = 0$ is a local minimum of the potential. Performing the Legendre transformation with respect to $\dot{\f}$ and $\dot{\ps}$, we obtain the Hamiltonian density
\beq
\label{ghostHamiltonianinteractingapp}
\mscr{H} = \frac{\ep}{2} \, \dot{\f}^2 + \frac{\ep}{2} \, \big( \vec{\nabla} \f \big)^{\! 2} + \frac{\ep}{2} \, m_{\f}^2 \f^2 + \frac{1}{2} \, \dot{\psi}^2 + \frac{1}{2} \, \big( \vec{\nabla} \psi \big)^{\! 2} + \frac{1}{2} \, m_{\psi}^2 \psi^2 + V_{\textup{int}} \big(\phi, \psi \big) \quad .
\eeq
Note first of all that, in the $V_{\textup{int}} = 0$ case, the state $\f = \ps = 0$ is still stable independently of the sign of $\ep$, as can be established performing an analysis analogous to the one performed in the single field case. However, this does \emph{not} happen because the only states which have energy close to $E = 0$ are small perturbations of the $\f = \ps = 0$ configuration. In fact, while this is true in the $\ep = +1$ case, if $\ep = -1$ there exist an infinite number of different configurations with $E = 0$ which are not small\footnote{We say that two configurations $f_1(\vecx)$ and $f_2(\vecx)$ are very close (respectively, very distant) if their difference $f_1 - f_2$ is small (respectively, big) in the sense of the previous footnote.} perturbations of the $\f = \ps = 0$ one (they are ``highly excited''). For example, the configurations where $\f$ and $\ps$ are plane waves with zero total 3-momentum and the same (arbitrary) amplitude, have zero total energy (where for simplicity we assumed $m_{\f} = m_{\ps}$, although this is not essential). The stability is instead due to the energy being separately conserved for the two fields, so the system cannot reach the infinite region in parameter space where both sectors are indiscriminately excited at fixed total energy.

If $V_{\textup{int}} \neq 0$, it remains true that configuration $\f(\vecx, t) = \ps(\vecx, t) = 0$ is a solution of the equations of motion, so if we prepare the system in the state $\f(\vecx, t_0) = \ps(\vecx, t_0) = 0$ at an initial time $t_0$, the fields $\f$ and $\ps$ remain in the ``vacuum'' configuration forever. To see what happens if we slightly perturb the vacuum configuration, note that the configurations introduced before, where the ghost and the ordinary field are plane waves of vanishing total momentum, now do not have zero energy any more due to the presence of the interaction terms. However, since derivative interactions are absent, choosing the amplitudes of the plane waves to be small enough we can construct configurations with energy as close to zero as we want, without constraining the wavevector of each plane wave whose magnitude can be arbitrarily big. Therefore, if $\ep = -1$, for every value of the energy $E \gtrsim 0$ there exists an infinite number of excited configurations and the volume of momentum space available for each (ordinary/ghost) sector is infinite (while this is not the case for $\ep = +1$). For entropy reasons, the decay towards these excited states is extremely favoured, and we conclude that the system is unstable to small oscillations.

Whether or not this perturbative instability develops to a finite amplitude instability, and the velocity with which this happens, depends both on the details of the interaction potential and on the initial conditions of the system (roughly speaking, the initial perturbation has to have enough power in the modes which are prone to the instability). On this respect, see the (toy model) analysis of \cite{Carroll:2003st} where a couple of a ghost harmonic oscillator and an ordinary harmonic oscillator (which are not 4D fields but simply functions of time) interact with a quartic potential $V_{\textup{int}} = \la \f^2 \ps^2$: the numerical simulations indicate that the system is non-linearly stable when the coupling constant $\la$ belongs to some range of values, while it is unstable in other cases. See also \cite{Woodard:2006nt} for a very interesting discussion about how the instability develops in higher derivative theories, and about the common misconceptions which are associated to it.

If the interaction Lagrangian contains derivative interactions, increasing the magnitude of the momentum in each sector may be costing from the energetic point of view, and the volume of momentum space available for decay into (small amplitude) plane waves may be finite. However, even if we construct the derivative interactions in such a way that high momenta for the ghost and the standard field are forbidden by energy conservation, there will always be a perturbative instability associated to the production of those plane waves configurations (with vanishing total 3-momentum) whose interaction energy is small. As we show in the next section, if the Lagrangian is Lorentz-invariant then the volume available for decay into small oscillations is infinite anyway, independently of the presence of ``stabilizing'' derivative interactions.

\section{Ghosts at quantum level}
\label{Ghosts at quantum level}

The presence of a ghost, already problematic at classical level, is even more so at quantum level. If we want to define the quantum theory of a field described by the Lagrangian density
\beq
\mscr{L}_{\f} = \frac{1}{2} \, \dem \f \, \de^{\m} \f + \frac{1}{2} \, m^2 \f^2 \quad ,
\eeq
we have two options: either the states which describe the quantum configuration of the field $\hat{\f}$ are assigned negative norm, or they are assigned positive norm (as usual in a quantum theory). The first choice implies that the energy spectrum is bounded from below, so the theory is stable, but the probabilistic interpretation of the theory is lost, and the theory is not predictive. To have a well-defined probabilistic interpretation, we have to choose the second option, which however implies that the energy spectrum is unbounded from below. From the point of view of the propagator, this ambiguity corresponds to the existence of two possible choices for the Feynman contour representation, or equivalently of two prescriptions for shifting the poles. For a ghost field, the first of the following momentum-space prescriptions
\begin{align}
&\frac{-i}{p^2 - m^2 + i \ep} & &\frac{-i}{p^2 - m^2 - i \ep}
\end{align}
causes the optical theorem to be violated and the loss of unitarity, while the second preserves unitarity but causes the particles with negative energy to propagate forward in time \cite{Cline:2003gs}. Unitarity then is granted at the price of having a theory which is prone to instabilities whenever the ghost field interacts with other (non-ghost) quantum fields. As we shall see, in this case the instability associated with the presence of the ghost field is much more severe at quantum level than it is at classical level.

Let's consider a relativistic ghost field $\f$ coupled to the Standard Model (SM) fields (collectively indicated with $\psi_{(j)}$) described schematically by the following local and Lorentz-invariant Lagrangian density
\beq
\label{quantumghostSMapp}
\mscr{L} = \mscr{L}_{\f}[\f,\de \f \,] + \mscr{L}_{\textup{SM}}[\psi_{(j)}, \de \psi_{(j)}] + \mscr{L}_{\textup{int}}[\f, \de \f ,\psi_{(j)}, \de \psi_{(j)}]
\eeq
where $\mscr{L}_{\textup{int}}$ describes the interaction of the ghost with the SM fields, while the self-interactions of the ghost and of the standard fields are contained respectively in $\mscr{L}_{\f}$ and in $\mscr{L}_{\textup{SM}}$. The ghost and the SM fields always couple at least gravitationally \cite{Cline:2003gs}, so there is always an effective interaction term, the interaction being graviton mediated, direct or both. Note that, if the ghost sector and the Standard Model sector were strongly coupled, the presence of the ghosts would be detected through fifth-force experiments or variation of the constants of nature \cite{Carroll:1998zi}. Moreover, if the interaction is purely graviton mediated we expect that the interaction terms between the ghost and the standard sector are suppressed by powers of the Planck mass. Therefore, we can assume that the interaction is weak at least at low energies. This implies that, if we consider a decay channel for the vacuum in which the final configuration $\mcal{F}$ is made of $n$ particles, the decay rate takes the form \cite{PeskinSchroeder}
\beq
\label{decayratenbodyapp}
\G_{\textup{vac} \rightarrow \mcal{F}} = \int \bigg( \prod_{i = 1}^{n} \frac{d^3 p_i}{(2 \pi)^3} \, \frac{1}{2 E_i} \bigg) \, \abs{\mcal{M}(\textrm{vac} \rightarrow \{ p_i \}_i)}^2 \, (2 \pi)^4 \, \d^{(4)} \bigg(\sum_{i = 1}^{n} p_i \bigg)
\eeq
where the $\{ p_i \}_i$ are the 4-momenta of the emitted particles and $p_i^0$ is the energy of the particle of mass $m_i$ and 3-momentum $\vecp_i$. The total decay rate is then the sum over all the possible decay channels
\beq
\G_{\textup{vac}} = \sum_{\mathcal{F}} \, \G_{\textup{vac} \rightarrow \mcal{F}} \quad .
\eeq
Strictly speaking, when $\mscr{L}_{\textup{int}} \neq 0$ and/or self-interactions are present, a configuration where each of the $n$ particles is described by a plane wave of definite 4-momenta $p_i$ is not an eigenstate of the Hamiltonian, and therefore is not an eligible final state for the decay of the vacuum (since the latter is an eigenstate of the Hamiltonian, and the energy is conserved by the time evolution). However, in the weak field approximation we can approximate the final states as products of single particle states of fixed momentum, and the total energy as the sum of the ``free'' energies of these states. Note that we are implicitly assuming that not only the interaction between the two sectors is weak, but also the self-interactions in each sector are weak. While this is an assumption for the ghost sector, it is justified for the standard sector since in most part of the universe the standard fields are weak. We comment on strong self-interactions of the ghost sector further on. The integral
\beq
\label{relinvnbodyphasespace}
\int \bigg( \prod_{i = 1}^{n} \frac{d^3 p_i}{(2 \pi)^3} \, \frac{1}{2 E_i} \bigg) \, (2 \pi)^4 \, \d^{(4)} \bigg(\sum_{i = 1}^{n} p_i \bigg)
\eeq
is called the \emph{relativistically invariant n-body phase space}, while $\mcal{M}(\textrm{vac} \rightarrow \{ p_i \}_i)$ is called the \emph{relativistically invariant transition matrix element}: in general, the decay rate depends both on the phase space available and on the modulation due to the dependence of the matrix element on the momenta.

Let's consider first a simplified case where a ghost field interacts with a standard field with a quartic coupling, so the interaction Lagrangian has the form $\mscr{L}_{\textup{int}} = \frac{\la}{4} \f^2 \psi^2$. For simplicity, we assume that both the ghost ($\f$) and the ordinary particle ($\ps$) coincide with their anti-particle. This interaction term allows a final state $\mathcal{F}$ which is a four-particle state made of a ghost-anti ghost couple and an ordinary particle-anti particle couple, since the transition matrix element for this final state is non-zero. Indicating with $\vecp_1$ and $\vecp_2$ the 3-momenta of the ghost particles, and with $\veck_1$ and $\veck_1$ the 3-momenta of the ordinary particles, the relativistically invariant 4-body phase space reads
\beq
\label{relinv4bodyphasespaceapp}
\mcal{I}_{\f\f\ps\ps} = \int \frac{d^3 p_1}{(2 \pi)^3 2 E_1} \, \frac{d^3 p_2}{(2 \pi)^3 2 E_2} \, \frac{d^3 k_1}{(2 \pi)^3 2 \o_1} \, \frac{d^3 k_2}{(2 \pi)^3 2 \o_2} \, \frac{(2 \pi)^4}{2! \, 2!} \, \d^{(4)} \big( p_1 + p_2 + k_1 + k_2 \big)
\eeq
where $p_i^0 = - \sqrt{m_{\f}^2 + \vecp_i^{\, 2}}$, $\o_i = \sqrt{m_{\ps}^2 + \veck_i^{\, 2}}$ and the factors $2!$ take into account that the particles $\f$ as well as the particles $\ps$ are identical. The minus sign in the expression for $p_i^0$ comes from the fact that, to preserve unitarity, for the ghost we have to choose the opposite contour representation of the Feynman propagator with respect to one for the healthy fields. The four dimensional delta function enforces the conservation of the total energy and momentum: it selects a volume $\mcal{V}$ in the 12-dimensional momentum space which contains the final momenta configurations which are compatible with the energy-momentum conservation. Note that, if $\f$ were an ordinary particle as well, then in the massless case only the state $\vecp_i = \veck_i = (0,0,0)$ would be compatible with the conservation of energy (and in the massive case no states at all): this implies that $\mcal{V}$ would have zero measure, and the vacuum would be stable (since $\G$ would vanish). The presence of quantum ghosts destabilize the vacuum because there exist ``excited'' states at the same energy of the vacuum, and so $\mcal{V}$ has non-zero measure: differently from the classical case, at quantum level we don't even need an initial perturbation to be able to reach these states, the vacuum decays spontaneously.

To evaluate the integral (\ref{relinv4bodyphasespaceapp}), we may integrate over $\veck_1$ and $\veck_2$ on the sections at fixed $\vecp_1$ and $\vecp_2$, and then integrate over $\vecp_1$ and $\vecp_2$. However, following \cite{Kaplan:2005rr,Jaccard:2013gla}, it is more convenient to embed $\mcal{V}$ into a 20-dimensional space, where the 8 extra dimensions are the components of the total ``ghost'' 4-momentum $P \equiv p_1 + p_2$ and the components of the total ``ordinary'' 4-momentum $K \equiv k_1 + k_2$, and calculate its area. We can in fact rewrite the total energy-momentum conservation as
\beq
\label{energyconservationdecomposition}
\d^{(4)} ( p_1 + p_2 + k_1 + k_2 ) = \d^{(4)} ( P + K ) \,\, \d^{(4)} ( P - p_1 - p_2 ) \,\, \d^{(4)} ( K - k_1 - k_2 )
\eeq
and then integrate over $\vecp_1$, $\vecp_2$ at fixed $P$, and independently integrate over $\veck_1$, $\veck_2$ at fixed $K$. The integration over $\vecp_1$, $\vecp_2$ generates the two-body phase space $\Phi_{\f}^{(2)}(-P^2)$ (which is defined by the general formula (\ref{relinvnbodyphasespace}) in the particular case where $n = 2$) for two identical particles of mass $m_\f$ whose center of mass energy is $- P^2$, while the integration over $\veck_1$, $\veck_2$ generates the two-body phase space $\Phi_{\ps}^{(2)}(-K^2)$ for two identical particles of mass $m_\ps$ whose center of mass energy\footnote{Note that, with our choice of signature for the metric, $-P^2$ and $-K^2$ are non-negative numbers.} is $- K^2$: we have then
\begin{align}
\mcal{I} &= \frac{1}{(2 \pi)^4} \int d^4 P \, d^4 K \, \d^{(4)} ( P + K ) \, \Phi_{\f}^{(2)}(-P^2) \, \Phi_{\ps}^{(2)}(-K^2) = \nn \\[2mm]
&= \frac{1}{(2 \pi)^4} \int d^4 P \, \Phi_{\f}^{(2)}(-P^2) \, \Phi_{\ps}^{(2)}(-P^2) \quad . \label{divergentapp}
\end{align}
The relativistically invariant two-body phase space for two identical particles of mass $m$ can be calculated explicitly, and reads \cite{Jaccard:2013gla}
\beq
\Phi^{(2)}(s) = \th(s - 4 m^2) \, \frac{1}{16 \pi} \, \sqrt{1 - \frac{4 m^2}{s}}
\eeq
where $\th(x)$ is the Heavyside theta function: it is easy to see that $\Phi^{(2)}(s)$ tends to a non-zero constant when $s \rightarrow + \infty$. Therefore, the integral (\ref{relinv4bodyphasespaceapp}) is badly divergent, as can be deduced from (\ref{divergentapp}). To characterize better the divergence, we can rewrite (\ref{divergentapp}) by integrating first over $P^0$ and $\vecP$ at $s = -P^2$ fixed, and then integrating the result over $s$. Adding a fifth dimension $s$ in the integration and inserting a delta function $\d(s + P^2)$, we get
\beq
\mcal{I}_{\f\f\ps\ps} = \frac{1}{(2 \pi)^4} \int_{0}^{+\infty} ds \, \Phi_{\f}^{(2)}(s) \, \Phi_{\ps}^{(2)}(s) \, \int_{\mathbb{R}^4} d P^0 \, d \vecP \,\, \d \big(s - (P^0)^2 + \vecP^2 \big)
\eeq
and using the property of the Dirac delta function
\beq
\d \big( f(x) \big) = \frac{1}{\abs{f^{\p}(x)}} \,\, \d(x)
\eeq
we obtain
\beq
\label{lecroquemonsieur}
\mcal{I}_{\f\f\ps\ps} = \frac{1}{(2 \pi)^4} \int_{0}^{+\infty} \!\! ds \, \Phi_{\f}^{(2)}(s) \, \Phi_{\ps}^{(2)}(s) \, \int_{\mathbb{R}^3} d \vecP \,\, \frac{1}{2 \sqrt{s + {\vec{P}}^{2}}} \quad .
\eeq
We note that not only the integral in $s$ is divergent, 
but the three-dimensional integral in $\vecP$ itself is divergent, since written in spherical coordinates it becomes
\beq
\label{messieursandmadames}
\int_{\mathbb{R}^3} d \vecP \,\, \frac{1}{2 \sqrt{s + {\vec{P}}^{2}}} = 2 \pi \int_{0}^{+\infty} \!\! d\z \, \frac{\z^2}{\sqrt{s + \z^2}}
\eeq
Therefore, for this type of interaction the decay rate of the vacuum to the emission of a ghost couple and a standard couple is divergent, which implies that the ``total'' decay rate $\G$ is divergent as well. In other words, the system is catastrophically unstable, since the mean lifetime $\t \sim 1/\G$ of the vacuum state vanishes. The same conclusion holds in the more general case where the ghost and the standard field does not coincide with their antiparticle. Note that the divergence is due to kinematics, i.e.~to the phase space available for the decay having infinite volume, and not to the actual value of the matrix element $\abs{\mcal{M}}^2$ (provided it is non-zero).

\subsection{Generality of the quantum instability}
\label{Generality of the quantum instability}

This result has wide implications. As we already mentioned, the gravitational interaction should induce at low energies a (graviton mediated) interaction between ghosts and standard fields. We expect quartic couplings of the type considered above to be generated this way, as well as a lot of other interaction terms: this is already enough to imply a diverging decay rate (always assuming Lorentz invariance and locality of the theory). 

Even if we neglect graviton-induced interactions and accept to consider an ad-hoc interaction Lagrangian, the conclusion does not change. Consider in fact a final state $\mcal{F} = \mcal{G} \otimes \mcal{S}$ where the ghost final configuration $\mcal{G}$ is made up of $n$ ghosts of definite 3-momenta $\{\vecp_i\}_{i = 1, \ldots , n}$ and the standard fields final configuration is made up of $n^{\p}$ fields with definite 3-momenta $\{\veck_i\}_{i = 1, \ldots , m}$. These factorized states constitute a basis of the Fock space for the system described by (\ref{quantumghostSMapp}), so for every form of $\mscr{L}_{\textup{int}}$ there exist some of these states such that the transition element $\mcal{M}$ is non-vanishing (at least in some range of values for the 3-momenta). It is implicitly assumed here that the final configurations $\mcal{S}$ span all the possible choices for $n^{\p}$ Standard Model fields which are allowed by the Standard Model selection rules. To evaluate the relativistically invariant phase space for this final state 
\beq
\label{relinvnbodyphasespaceGS}
\mcal{I}_{\mcal{G} \otimes \mcal{S}} = \int \bigg( \prod_{i = 1}^{n} \frac{d^3 p_i}{(2 \pi)^3} \, \frac{1}{2 E_i} \, \prod_{i = 1}^{n^{\p}} \frac{d^3 k_i}{(2 \pi)^3} \, \frac{1}{2 \omega_i} \bigg) \, (2 \pi)^4 \, \d^{(4)} \bigg(\sum_{i = 1}^{n} p_i + \sum_{i = 1}^{n^{\p}} k_i \bigg)
\eeq
we can follow the same procedure used above. Indicating with $P$ and $K$ respectively the total 4-momentum of the ghost particles and of the standard particles, and using the appropriate generalization of the relation (\ref{energyconservationdecomposition}), we get 
\beq
\mcal{I}_{\mcal{G} \otimes \mcal{S}} = \frac{1}{(2 \pi)^4} \int d^4 P \, \Phi_{\mcal{G}}^{(n)}(-P^2) \, \Phi_{\mcal{S}}^{(n^{\p})}(-P^2) 
\eeq
where $\Phi_{\mcal{G}}^{(n)}$ and $\Phi_{\mcal{S}}^{(n^{\p})}$ are respectively the relativistically invariant $n$-body ($n^{\p}$-body) phase space for the ghost sector (for the standard sector). Inserting also in this case a delta function $\d(s + P^2)$ we arrive at
\beq
\label{voila}
\mcal{I}_{\mcal{G} \otimes \mcal{S}} = \frac{1}{(2 \pi)^4} \int_{0}^{+\infty} \!\! ds \, \Phi_{\mcal{G}}^{(n)}(s) \, \Phi_{\mcal{S}}^{(n^{\p})}(s) \, \int_{\mathbb{R}^3} d \vecP \,\, \frac{1}{2 \sqrt{s + {\vec{P}}^{2}}} \quad .
\eeq
This $(n+n^{\p}$)-body phase space is divergent because of the integral in $\vecP$, independently of the behaviour of $\Phi_{\mcal{G}}^{(n)}(s)$ and $\Phi_{\mcal{S}}^{(n^{\p})}(s)$. The crucial point is that, since the theory is Lorentz-invariant, $\Phi_{\mcal{G}}^{(n)}$, $\Phi_{\mcal{S}}^{(n^{\p})}$ and $\mcal{M}$ can depend on $P$ only through $s$, while are independent of $\vecP$: therefore cannot modulate the integral in $\vecP$ and render it convergent.

The catastrophic instability of the vacuum is therefore a general consequence of the fact that the theory is local and Lorentz-invariant, and that the coupling between the ghost and the standard sector is weak at low energies. Note that this conclusion is independent of the presence of (Lorentz-invariant) derivative interactions. Even if we consider ad-hoc derivative interactions which render the Hamiltonian bounded from below (for example appropriate self-interactions of the ghost sector which become strong at a certain scale), their effect is merely to cut-off the integration is $s$ but cannot influence the integration in $\vecP$.

\section{Ghosts in ef\mbox{}fective theories}
\label{Ghosts in effective theories}

It is worthwhile to spend few words on the physical meaning of the decay rate of the vacuum. The decay rate is a probability density (in space and time) of decay. Focusing (to fix ideas) on the decay channel $\textrm{vac} \rightarrow \f\f \g\g$ , if $V$ is a 3-dimensional volume and $T$ is a time interval then the quantity
\beq
N_{\f \f \g \g} = V \, T \, \G_{vac \rightarrow \f \f \g \g}
\eeq
gives the average number of quadruplets $\f \f \g \g$ emitted in the volume $V$ in the time interval $T$ by the decay of the vacuum (regardless of the momenta of the emitted particles). If in the integral (\ref{decayratenbodyapp}) we integrate just over a specific interval of energy for the emitted particles, the result multiplied by $V$ and $T$ gives the average number of quadruplets $\f \f \g \g$ emitted in the volume $V$ in the time interval $T$ which have energies in the selected interval. In an expanding universe, we need to take into account the effect of the expansion on their density and on their spectrum. The results of the previous section imply that we expect to detect an infinite number of ghosts and photons emitted by the vacuum decay in every spatial volume and time interval, no matter how small. Furthermore, this remains true even if we focus only on any specific energy interval for the photons, since we get a diverging result even if in (\ref{lecroquemonsieur}) we integrate just over a finite integral in $s$. This is true also for the spontaneous emission of electron-positron couples $\textrm{vac} \rightarrow \f\f \, e^{-} e^{+}$, neutrino-antineutrino couples $\textrm{vac} \rightarrow \f\f \, \n \bar{\n}$ and in general for every allowed decay channel.

This prediction is clearly in contrast with observations. It seems therefore that the possibility of considering systems described by a Lagrangian density of the form (\ref{quantumghostSMapp}) is ruled out observationally. However, this conclusion follows from the fact that the domain of integration over $s$ and $\vecP$ in (\ref{voila}) is not bounded, i.e.~from the assumption that the Lagrangian density (\ref{quantumghostSMapp}) is the correct description of nature at arbitrarily high energies and arbitrarily small distances. If instead we assume that such a Lagrangian density is \emph{not} an exact description of reality, but is rather an effective description which can be trusted only in a definite range of energy/length scales, then this conclusion may change. As a matter of fact, it is widely assumed that the Standard Model itself can be trusted only below an energy cut-off which is at least at the TeV scale, where it has been tested experimentally. Concerning the full Lagrangian, it may be that at high energies the coupling between the ghost sector and the standard sector is negligible, or is not local and Lorentz-invariant. Another possibility is that the very existence of the ghost sector is just a low energy effective property, and that the fundamental theory is ghost free. In this case, by definition the coupling between the two sectors vanishes at high energies.

Nevertheless, the results of the previous section imply that, to have a finite decay rate, we need to cut-off both the integration over $s$ \emph{and} the integration over $\vecP$. This seems impossible to achieve if the correct description of nature above the cut-off is still Lorentz-invariant. In fact, in this case the cut-off itself has to be imposed in a Lorentz-invariant way, independently on the nature of the UV completion of the action (\ref{quantumghostSMapp}). The only Lorentz-invariant way to put a cut-off on the total ghost momentum $P$ is to restrict the domain of $s = P^2$, without affecting the integration over $\vecP$ at $s$ fixed.

It is important to point out that some authors have assumed that a ghost in an effective theory can be harmless, as long as its mass is high enough. This is however not true since, for what concerns the instability, it is irrelevant if the mass of the ghost is higher or lower than the cut-off of the effective theory. In fact, a positive energy mode in a stable effective theory decouples when its mass is higher than the cut-off, because exciting it costs an amount of energy which is at least equal to its mass. However, no energy is needed to excite a negative energy mode, since the decay happens through simultaneous emission of ghost modes and healthy modes at fixed total energy. Therefore a ghost does not decouple when its mass is higher than the cut-off, actually it couples more strongly since (from the point of view of the energy) it is more effective in exciting the healthy modes \cite{Woodard:2006nt,Woodard:2014private}.

\subsection{Spontaneous breaking of Lorentz symmetry}

It is however important to take into account that Lorentz symmetry is spontaneously broken in our universe, since the latter appears homogeneous and isotropic (on large scales) only in a class of reference systems which is not closed under Lorentz transformations. In particular the Cosmic Microwave Background radiation (CMB) singles out a class of preferred frames. If we want to be maximally conservative, in the integration (\ref{decayratenbodyapp}) we should not consider processes with a formation time longer than the age of the universe \cite{Cline:2003gs,Kaplan:2005rr,Garriga:2012pk}. 
This is a Lorentz-violating condition and effectively cuts-off the integration in $\vecP$. It is important to point out that the existence of such a condition is not actually dependent on our universe being non-eternal. In general, the vacuum of a system containing ghosts can be at best metastable, since its decay rate is non-vanishing (if not diverging). If such a vacuum were created at $t \rightarrow -\infty$, it would have already decayed anyway, producing a infinite amount of radiation (independently of the actual value of its decay rate). This implies that, in an eternal universe, a vacuum with ghosts has to be created at a certain point on some space-like hypersurface $\S$. In an eternal inflation scenario, for example, the ghost-carrying vacuum could be created inside certain kind of bubbles from an earlier ghost-free vacuum: in this case, the role of $\S$ would be played by the hypersurface where the phase transition occurs. We conclude that Lorentz-invariance necessarily has to be broken spontaneously if we consider vacuum states with ghosts, and the presence of the space-like hypersurface $\S$ (be it the ``comoving'' $t = 0$ hypersurface in standard cosmology or something more exotic) introduces an effective cut-off on the $\vecP$ integration (\ref{messieursandmadames}).

\subsubsection{Experimental bound and derivative self-interactions}
\label{Experimental bound and derivative self-interactions}

The cut-off on the integration over $\vecP$ due to the finite age of the universe suggests that it may be possible for a vacuum with ghosts to have a finite decay rate without giving up locality and Lorentz-invariance of the physical laws. In fact, if the action (\ref{quantumghostSMapp}) is an effective description valid below a Lorentz-invariant cut-off $\La$, the combined action of the two cut-offs (the one induced by the spontaneous Lorentz symmetry breaking and the Lorentz-invariant one) can render the integral (\ref{voila}) convergent. It is important to estimate how low the cut-off on $s$ has to be to produce a decay rate compatible with observations. As it is explained in \cite{Cline:2003gs}, the most stringent bound on $\La$ is found by considering the decay channel $\textrm{vac} \rightarrow \f\f\g\g$: the most energetic of the emitted photons would scatter with CMB photons according to $\g\g \rightarrow e^{-}e^{+}e^{-}e^{+}$ and would produce air showers of particles when arriving at the Earth. Compatibility with the observations then implies that
\beq
\label{bound1}
\La \lesssim 10^{-3} \, \textrm{eV}
\eeq
while the other decay channels (for example $\textrm{vac} \rightarrow \f\f e^{-}e^{+}$ and $\textrm{vac} \rightarrow \f\f \n \bar{\n}$) produce less restrictive bounds \cite{Cline:2003gs}. This cut-off is much lower than the cut-off of the Standard Model, which is well tested (at least) till the TeV scale. This implies that the breakdown of the description (\ref{quantumghostSMapp}) at $\La$ cannot be ascribed to the standard sector.

In principle it may be possible to construct a local and Lorentz-invariant Lagrangian for the (derivative) self-interaction of the ghost sector, such that it renders the Hamiltonian bounded from below and becomes strongly interacting at the meV scale. This would effectively cut-off the integration over $s$ at the scale indicated in (\ref{bound1}). From another point of view, we may postulate that the existence of the ghosts is a low-energy effective property and they are not present above the meV scale. Despite being in principle feasible, these two possibilities necessitate of a very ad-hoc and fine-tuned structure of the theory, and a priori seem artificial. Furthermore, it is not clear how to construct explicitly a ghost-free theory which reproduces the Standard Model below the TeV scale, and such that effective ghosts appear at even lower energies. A more natural possibility would be that the two sectors exist also above the meV scale, but are completely decoupled above the cut-off (\ref{bound1}). This is however experimentally excluded. Since the graviton-mediated interaction between the ghosts and the standard particles is unavoidable in GR, for this idea to work we need to postulate that GR is not the correct description of gravity above the meV scale. But this would imply that gravity departs from the GR predictions at a length scale $> 0.2$ mm \cite{Cline:2003gs}, which is experimentally excluded \cite{Kapner:2006si}. Therefore, if we assume that locality and Lorentz-invariance are fundamental properties, the cut-off on the integration over $\vecP$ due to the finite age of the universe is not able to render the decay rate of the vacuum compatible with observations, unless we invoke very ad-hoc choices.

\subsection{Non-locality and Lorentz violation}

The results above imply that, apart from artificial constructions, an effective theory with perturbative ghosts can be phenomenologically viable only if the correct description of nature is either non-local or Lorentz-violating above the cut-off, or both.

\subsubsection{Non-locality}

Interestingly, there is a claim \cite{Garriga:2012pk} that if the theory describing the interaction of the ghost and the standard sector remains Lorentz-invariant but becomes \emph{non-local} above a (Lorentz-invariant) cut-off, the decay rate of the vacuum can be compatible with observations. As an example, consider a ghost field $\f$ and a standard field $\ps$ whose interaction is described by the action
\begin{multline}
\label{non local interaction}
S_{I} = \frac{\la}{4} \int d^{4} x \, d^{4} z \, d^{4} y_{1} \, d^{4} y_{2} \,\, \f \big(x +z +y_1 \big) \, \f \big(x +z - y_1 \big) \\
g\big(z, y_1, y_2 \big) \, \ps \big(x -z +y_2 \big) \, \ps \big(x -z - y_2 \big) \quad ,
\end{multline}
which is the non-local generalization of a quartic $\la \, \f^2 \ps^2$ coupling. Here $g(z, y_1, y_2)$ is the non-local form factor, while $y_1$ (respectively, $y_2$) is the coordinate distance between the points at which the two ghosts (respectively, standard) fields interact, and $z$ is the coordinate distance between the interaction points of the ghost couple and the standard couple. The non-local properties of the interaction, and in particular the fact that the interaction is non-local above or below a cut-off, are linked to the properties of the Fourier transform $G(q^{\m}, q_1^{\m}, q_2^{\m})$ of the form factor, where $q^{\m}$, $q_1^{\m}$ and $q_2^{\m}$ are the 4-momenta dual to the coordinates $z$, $y_1$ and $y_2$. In particular, the interaction is local when $G$ is constant, and so $g$ is 12-dimensional Dirac delta. If the theory is Lorentz-invariant, then $G$ can only depend on the square moduli of $q^{\m}$, $q_1^{\m}$ and $q_2^{\m}$ and on the scalar products $\pi_1 \equiv q_\m \, q_1^{\m}$, $\pi_2 \equiv q_\m \, q_2^{\m}$ and $\pi_{12} \equiv q_{1 \m} \, q_2^{\m}$. A Lorentz-invariant cut-off on the theory constrains the values of these 6 Lorentz-invariant quantities, and restricts the domain of integration in momentum space involved in the calculation of the decay amplitude. The authors of \cite{Garriga:2012pk} claim that, if the theory (\ref{non local interaction}) is Lorentz-invariant and non-local above a cut-off $\La$, the Lorentz-violating cut-off due to the finite age of the universe can be sufficient to produce a decay rate consistent with the observations, since the Lorentz-preserving cut-off can be slightly higher than the previous bound
\beq
\La \lesssim (1.8 - 5.6) \times 10^{-3} \, \textrm{eV} \quad ,
\eeq
which is marginally consistent with the experimental data on small distances modifications of GR \cite{Garriga:2012pk,Kapner:2006si}. Note that non-locality does not necessarily imply a lack of causality in the theory \cite{Jaccard:2013gla}, and that, even though a generic non-local theory may violate Lorentz-invariance, there exists a class of non-local theories which does not violate it \cite{Garriga:2012pk}. In this case, the breakdown of GR and the appearance of signals of new gravitational physics would be very close to the present experimental reach.

\subsubsection{Lorentz violation}

If we demand locality of interactions to be a fundamental property of nature, then (with the proviso of section \ref{Experimental bound and derivative self-interactions}) the only way to accommodate perturbative ghosts in a low-energy effective description is to assume that the theory above the cut-off is Lorentz-violating. In fact, in this case the cut-off on the integration over $\vecP$ due to the finite age of the universe is not enough to render the decay rate compatible with observations, which implies that the Lorentz-violating cut-off has to be due to the structure of the theory itself.

The most stringent bound on the (Lorentz-violating) energy cut-off $\La$ comes again from the decay channel $\textrm{vac} \rightarrow \f\f\g\g$ of the vacuum, and the observations on the diffuse gamma ray background imply \cite{Cline:2003gs}
\beq
\label{bound2}
\La \lesssim 3 \, \textrm{MeV} \quad .
\eeq
This is still lower than the cut-off of the Standard Model, although higher than (\ref{bound1}). If we assume that the ghost sector exists also above the cut-off (\ref{bound2}), then (analogously to the discussion in section \ref{Experimental bound and derivative self-interactions}) this implies that GR cannot be the correct description of gravity above the cut-off, where the ghosts and the ordinary fields have to be completely decoupled. Differently from the discussion in section \ref{Experimental bound and derivative self-interactions}, this is not excluded experimentally, since in this case deviations of gravity from the GR predictions at small distances would happen well below the present experimental reach. Most importantly, in this case Lorentz-invariance itself, and not only GR, has to break down well below the TeV scale. Despite there are very severe constraints on Lorentz-violation within ordinary particle physics, and Lorentz-violation in another sector tends to be communicated to the standard sector via graviton loops \cite{Cline:2003gs}, these effects are expected to be negligibly small \cite{Kaplan:2005rr}.

We arrive at the same conclusions if we assume that the ghost sector does not exist above the cut-off (\ref{bound2}). In fact, while in this case it is not necessary to get rid of gravitational couplings above the cut-off, Lorentz-invariance has still to be violated above the scale (\ref{bound2}). Since Lorentz-invariance (full diffeomorphism invariance, in fact) is at the core of General Relativity, we conclude that the latter has to break down well below the TeV scale.

\section{Discussion}

The results above imply that, if we avoid very ad-hoc choices, we could accommodate perturbative ghosts in a (local and Lorentz-invariant) effective description only if locality and/or Lorentz-invariance are/is violated above the cut-off. This means that, if we accept the presence of effective perturbative ghosts, we have to assume that locality and/or Lorentz-invariance are accidental or emerging properties, and not fundamental properties of the correct description of nature. Furthermore, the breakdown of these effective properties has to happen well \emph{below} the energies we can probe in particle colliders. Despite this is not experimentally excluded, it is a very unorthodox situation. This conclusion can be relaxed postulating that, alongside the sector of Standard Model particles and the ghost sector, there exist extra degrees of freedom whose configuration breaks spontaneously Lorentz-invariance \cite{Kaplan:2005rr}. This possibility was not considered above, since we focused on an effective Lagrangian of the type (\ref{quantumghostSMapp}), but it may be attractive from the point of view of the cosmological late time acceleration problem (see for example the effective field theory of ref.~\cite{ArkaniHamed:2003uy}).

There is a subtlety underlying the analysis of sections \ref{Ghosts at quantum level} and \ref{Ghosts in effective theories}, namely that the quantization is semiclassical. In fact, we did not quantize the theory of the fields $\{ \Psi_i \}_{i = 1, \ldots, N}$, but we considered a solution $\{ \bar{\Psi}_i \}_{i = 1, \ldots, N}$ of the classical equations of motion and quantized only the perturbations around the classical background. This is however justified when we consider (\ref{quantumghostSMapp}) as an effective Lagrangian. 
It is also important to point out that we implicitly assumed throughout the paper that the perturbative ghost modes are independent degrees of freedom (in a Hamiltonian sense). If this is not true, then the quantization is not performed correctly, since before quantizing we should identify the true degrees of freedom of the theory. We finally note that, despite we showed that a configuration with perturbative ghosts is always unstable, this is not necessarily true for any configuration of a theory with (fundamental) ghosts. In fact, a theory where (in our language) some of the fields $\{ \Psi_i \}_{i = 1, \ldots, N}$ are ghosts can admit solutions $\{ \bar{\Psi}_i \}_{i = 1, \ldots, N}$ of the equations of motion such that the perturbation modes around this background are ghost-free \cite{ArkaniHamed:2003uy}.

\section*{Acknowledgements}
The author wishes to thank Antonio Padilla for encouraging him to render these notes public. The author also acknowledges useful conversations with Kazuya Koyama and Gianmassimo Tasinato, and a clarifying email exchange with Richard Woodard.

\end{document}